\documentclass[preprint2,twocolumn,times,tighten]{aastex631}
\usepackage{graphicx}
\usepackage{amsmath}
\usepackage{amssymb}
\let\tablenum\relax
\usepackage{siunitx}

\newcommand{\nh}{n_{\rm H}}
\newcommand{\Ms}{\mathcal{M}_{\rm s}}
\newcommand{\Ma}{\mathcal{M}_{\rm A}}
\newcommand{\kms}{\mathrm{km\,s^{-1}}}

\newcommand{\Fflat}{\mathcal{F}}

\begin{document}

\title{Intermittency in the Multiphase ISM --- I. Extreme Density, Velocity, and Magnetic Fluctuations}

\email{yuehu@caltech.edu}

\author[0000-0002-8455-0805]{Yue Hu}
\affiliation{Cahill Center for Astronomy and Astrophysics, California Institute of Technology, Pasadena, CA 91125, USA}

\begin{abstract}
Intermittency is a fundamental property of astrophysical turbulence because it concentrates strong gradients, dissipation, and transport into rare, localized structures. In the multiphase interstellar medium (ISM), turbulent driving is especially important because the balance between compressive and solenoidal motions can regulate how extreme fluctuations are distributed among thermal phases. Yet conventional increment statistics quantify intermittency without identifying the structures that host it. We analyze three matched ideal-MHD simulations of a thermally multiphase neutral medium that differ only in the compressive fraction of the driving. Density, velocity, and magnetic-field increments are non-Gaussian in every thermal phase. Increasing compressive driving redistributes density and magnetic fluctuations preferentially into the cold gas rather than uniformly enhancing them across phases. We identify the structures associated with extreme increments using local, scale-matched diagnostics of compression, thermal gradients, vorticity, strain, magnetic curl, and magnetic tension. Extreme magnetic increments preferentially occur where magnetic curl and magnetic tension are simultaneously enhanced, more strongly than extreme velocity increments are associated with their kinematic hosts. Extreme velocity increments show comparable enrichment in vorticity and strain, whereas extreme density increments preferentially trace converging flows and thermal gradients only in the unstable medium. These associations weaken beyond a phase-dependent coherence scale. Our results show that turbulent driving controls how intermittency is partitioned among thermal phases, while distinct physical structures host extreme density, velocity, and magnetic fluctuations. 
\end{abstract}

\keywords{Interstellar medium (847) --- Interstellar dynamics (839) ---
Magnetohydrodynamics (1964)}

\section{Introduction}\label{sec:introduction}

The diffuse neutral interstellar medium (ISM) plays a central role in galactic evolution: it supplies the material from which molecular clouds and stars form \citep{1977ApJ...218..148M,2004RvMP...76..125M,2007ApJ...657..870V,2009Natur.457...63G,2012ApJ...761..156F,2018ApJ...863..118B,2019FrASS...6....7K,2020ApJ...905..129H}, mediates the transport of magnetic flux and turbulent energy \citep{1995ApJ...443..209A,2010ApJ...710..853C,2013MNRAS.436.1245F,2019ApJ...878..157X,2022ApJ...934....7H,2024MNRAS.527.3945H,2025ApJ...986...62H,2026arXiv260808448H}, and shapes Galactic dust emission, a dominant foreground for CMB-polarization measurements \citep{2008MNRAS.388..117H,2015ARA&A..53..501A,2020A&A...641A..11P,2021ApJ...909...94D,2026arXiv260117255H}. It is simultaneously turbulent, magnetized, and thermally multiphase. Radiative heating and cooling produce cold and warm stable branches separated by a thermally unstable regime \citep{1965ApJ...142..531F,2003ApJ...587..278W}; turbulent compression, rarefaction, and mixing continually move gas among them \citep{2000ApJ...540..271V,2002ApJ...564L..97K,2011A&A...526A..14S,2014A&A...567A..16S,2021MNRAS.505.5238J,2024arXiv240714199H,2025ApJ...986...62H}. A single sound speed or isothermal density distribution is therefore insufficient to characterize the flow.

Intermittency arises because turbulent energy and extreme fluctuations are not distributed uniformly throughout the ISM, but are concentrated within a small fraction of its volume \citep{1993PhRvE..48...29B,1994PhRvL..72..336S,2009A&A...500L..29H}. In MHD turbulence, modern theories connect intermittency to aligned, sheet-like magnetic structures and their disruption \citep{2007ApJ...658..423K,2022JPlPh..88e1501S}. In the ISM, localized velocity shears and dissipation have been linked to chemical heating and small-scale structure \citep{2008A&A...481..367H,2009A&A...500L..29H,2009A&A...495..847G,2021MNRAS.500.3290M}; recent molecular-cloud measurements likewise find intermittent, non-Gaussian velocity statistics \citep{2025ApJ...979...65M}. Increment PDFs, high-order structure functions, and extended self-similarity quantify the scale dependence \citep{1993PhRvE..48...29B,1994PhRvL..72..336S}, but do not identify the physical hosts. A structural description is needed to determine whether extreme density, velocity, and magnetic increments occupy the same regions of the flow.

Two influences are expected to shape this structural organization in the neutral ISM. The first is the turbulent driving. Solenoidal driving is associated with shear and vortical motions, whereas compressive driving arises more directly from shocks and converging flows; relative to solenoidal forcing, compressive forcing produces broader density distributions and substantially larger density contrasts in otherwise comparable isothermal simulations \citep{2008ApJ...688L..79F,2010A&A...512A..81F}. The actual ISM is driven by a mixture of both, so the response to the degree of compressive forcing---and not only the contrast between its two limits---is the quantity of physical interest. The second is thermal phase separation, since the cold, unstable, and warm components occupy distinct dynamical regimes, with different sound speeds, plasma $\beta$, and local sonic and Alfv\'enic Mach numbers, within the same turbulent flow \citep{2021NatAs...5..365F,2026arXiv260117255H}. These influences have been studied separately: controlled forcing comparisons usually assume isothermal turbulence \citep{2008ApJ...688L..79F,2010A&A...512A..81F}, whereas work on thermally multi-phase turbulence has focused on phase morphology, energetics, and low-order statistics \citep{2011A&A...526A..14S,2021MNRAS.505.5238J,2025ApJ...986...62H}. Cloud velocity statistics also change with gravity, magnetic fields, radiation, and feedback \citep{2022MNRAS.513.2100H}, motivating a controlled multiphase baseline. In addition, each thermal phase occupies finite structures, so phase-conditioned statistics necessarily change character once the separation exceeds their characteristic size.

In this work, we characterize the intermittency of the diffuse neutral ISM and identify the structures that carry it, using three ideal-MHD simulations of a thermally multiphase, $100$~pc medium at a resolution of $1024^3$, plus a $2048^3$ solenoidally driven realization for a targeted resolution test. The three forcing runs share the same physical and numerical parameters and differ only in the compressive fraction of the driving, allowing us to trace the response to the driving rather than only bracket it. We decompose each run into cold, unstable, and warm components. We first characterize intermittency through the scale-dependent increments and flatness of density, velocity, and magnetic field. We then measure, for each field, the local environment in which the most extreme increments occur---compression, vorticity, shear, current density, magnetic-field
curvature, or a thermal interface---and determine how this environment depends on the thermal phase, on the driving, and on the scale at which the increments are taken.

This paper is organized as follows. In Section~\ref{sec:methods} we describe the simulations, the increment statistics, and the conditional diagnostics used to locate the extreme fluctuations. In Section~\ref{sec:results} we present the scale-dependent intermittency of the density, velocity, and magnetic fields in the full medium and in its individual thermal components, the dependence on the driving, and the structural hosts of the extreme increments. We discuss the physical implications in
Section~\ref{sec:discussion} and summarize our conclusions in Section~\ref{sec:conclusions}.

\begin{deluxetable*}{ccccccccccc}

\tablecaption{Global and phase-conditioned properties of the simulations.
\label{tab:sim-properties}}

\tablehead{
\colhead{Forcing} &
\colhead{Phase} &
\colhead{$\langle n_{\rm H}\rangle$ (cm$^{-3}$)} &
\colhead{$\sigma_{\ln n}$} &
\colhead{$|\mathbf{v}|$ (km s$^{-1}$)} &
\colhead{$|\mathbf{B}|$ (\SI{}{\micro G})} &
\colhead{med($\mathcal{M}_s$)} &
\colhead{med($\mathcal{M}_A$)} &
\colhead{med($\beta$)} &
\colhead{$f_V$} &
\colhead{$f_M$}
}

\startdata
Compressive & Total & $1.0$ & $1.3$ & $9.1$ & $3.1$ & $1.2$ & $0.9$ & $0.7$ & $1.00$ & $1.00$ \\
Compressive & WNM & $0.3$ & $1.0$ & $9.2$ & $3.0$ & $1.1$ & $0.8$ & $0.7$ & $0.59$ & $0.19$ \\
Compressive & UNM & $1.0$ & $1.3$ & $9.1$ & $3.1$ & $1.5$ & $1.3$ & $0.8$ & $0.40$ & $0.38$ \\
Compressive & CNM & $35.2$ & $0.8$ & $8.1$ & $9.5$ & $7.4$ & $2.5$ & $0.1$ & $0.01$ & $0.43$ \\ \hline
Mixed & Total & $1.0$ & $0.9$ & $9.6$ & $4.5$ & $1.3$ & $0.8$ & $0.4$ & $1.00$ & $1.00$ \\
Mixed & WNM & $0.5$ & $0.7$ & $9.6$ & $4.4$ & $1.1$ & $0.6$ & $0.4$ & $0.61$ & $0.28$\\
Mixed & UNM & $1.3$ & $0.8$ & $9.7$ & $4.6$ & $1.6$ & $1.0$ & $0.4$ & $0.39$ & $0.51$ \\
Mixed & CNM & $25.7$ & $0.6$ & $9.3$ & $6.4$ & $8.0$ & $3.3$ & $0.2$ & $0.01$ & $0.22$ \\ \hline
Solenoidal & Total & $1.0$ & $0.8$ & $10.0$ & $4.7$ & $1.3$ & $0.8$ & $0.4$ & $1.00$ & $1.00$ \\
Solenoidal & WNM & $0.5$ & $0.6$ & $9.8$ & $4.7$ & $1.2$ & $0.7$ & $0.4$ & $0.59$ & $0.31$ \\
Solenoidal & UNM & $1.3$ & $0.7$ & $10.1$ & $4.7$ & $1.7$ & $1.0$ & $0.4$ & $0.40$ & $0.53$ \\
Solenoidal & CNM & $25.2$ & $0.6$ & $9.9$ & $6.0$ & $8.5$ & $3.8$ & $0.2$ & $0.01$ & $0.17$ \\
\enddata
\tablecomments{Here $\sigma_{\ln n}\equiv \sigma_{\ln(n_{\rm H}/\langle n_{\rm H}\rangle)}$ within each mask. The quantities $|\mathbf{v}|$ and $|\mathbf{B}|$ are volume means of the velocity magnitude and magnetic-field strength, respectively. The quantities $\mathcal{M}_s$, $\mathcal{M}_A$, and $\beta$ are medians of the local cell values, using the adiabatic sound speed $c_s$ and Alfv\'en speed $v_A$ as defined in the phase-diagram analysis. The Total rows are domain-wide, while the WNM, UNM, and CNM rows give the corresponding phase-conditioned quantities. The final two columns list the volume and mass fractions, $f_V$ and $f_M$, respectively. Each entry is the mean over the snapshots.}
\end{deluxetable*}

\section{Numerical and analysis methods}\label{sec:methods}

\subsection{Multiphase-ISM simulations}\label{sec:simulation}
We solve the equations of ideal compressible MHD in a triply periodic cube with the \texttt{AthenaK} code \citep{2025ApJ...986...62H,2026ApJS..283...27S},
\begin{align}
 \frac{\partial \rho}{\partial t}+\boldsymbol{\nabla}\cdot(\rho\boldsymbol{v})&=0,
 \label{eq:cont}\\
 \frac{\partial(\rho\boldsymbol{v})}{\partial t}
 +\boldsymbol{\nabla}\cdot\left[
 \rho\boldsymbol{v}\boldsymbol{v}
 +\left(P+\frac{B^2}{8\pi}\right)\mathbf{I}
 -\frac{\boldsymbol{B}\boldsymbol{B}}{4\pi}\right]
 &=\rho\boldsymbol{f},
 \label{eq:mom}\\
 \frac{\partial\boldsymbol{B}}{\partial t}
 -\boldsymbol{\nabla}\times(\boldsymbol{v}\times\boldsymbol{B})&=0,
 \label{eq:induction}\\
 \nabla\cdot \boldsymbol{B} &= 0,
\end{align}
where $\rho$ is the mass density, $\boldsymbol{v}$ the velocity, $P$ the thermal pressure, $\boldsymbol{B}$ the magnetic field, and $\boldsymbol{f}$ the driving acceleration. These are supplemented by an energy equation with net volumetric cooling
\begin{equation}
\mathcal{L}=\nh^2\Lambda(T)-\nh\Gamma,
\label{eq:cooling}
\end{equation}
where $\nh$ is the hydrogen number density and $T$ the temperature. The cooling rate $\Lambda$ accounts for atomic line cooling and is taken from
\citet{2002ApJ...564L..97K}:
\begin{equation}
\begin{aligned}
\Lambda = \left(\frac{\rho}{m_{\rm H}}\right)^2 & \left[2\times10^{-19}\exp\left(\frac{-1.148\times 10^5}{T+1000}\right) \right. \\
& \left. + 2.8\times10^{-28}\sqrt{T}\exp\left(\frac{-92}{T}\right)\right] ~~ \mathrm{erg~s^{-1}~cm^{-3}},
\end{aligned}
\end{equation}
where $m_{\rm H}$ is the hydrogen mass, and the photoelectric heating rate $\Gamma$ is
\begin{equation}
\Gamma = \left(\frac{2\rho}{m_{\rm H}}\right)\times10^{-26} ~~ \mathrm{erg~s^{-1}~cm^{-3}}.
\end{equation}

The box size is \(L_{\rm box}=100\)~pc and the uniform mesh contains \(1024^3\) cells, giving an effective numerical dissipation scale of approximately ten cells. We neglect thermal conduction because its characteristic length scale ($\sim 10^{-2}$~pc) is not resolved at the adopted resolution. The initial mean hydrogen number density is \(n_{\rm H,0}=1~\mathrm{cm^{-3}}\) and the initial magnetic field is uniform with $B_0 \approx 3$~\SI{}{\micro G}, consistent with Zeeman observations \citep{2012ARA&A..50...29C}. The velocity dispersion reaches $\sigma_v \approx 10$~km~s$^{-1}$, based on Larson's law and on observational constraints derived from young stellar environments \citep{1981MNRAS.194..809L,2022ApJ...934....7H}. 

We drive the turbulence on scales from $L_{\rm inj}=100$ to $50$ pc. The three simulations use identical numerical, thermodynamic, and magnetic setups and differ only in the mixture of solenoidal and compressive forcing. The driving acceleration is composed of a solenoidal part \(\boldsymbol{f}_{\rm sol}\), with \(\boldsymbol{\nabla}\cdot\boldsymbol{f}_{\rm sol}=0\), and a compressive part \(\boldsymbol{f}_{\rm comp}\), with \(\boldsymbol{\nabla}\times\boldsymbol{f}_{\rm comp}=0\). The parameter \(f_{\rm sol}\) sets the weight of the two parts,
\begin{equation}
  \boldsymbol{f} = f_{\rm sol}\,\boldsymbol{f}_{\rm sol}
  + (1-f_{\rm sol})\,\boldsymbol{f}_{\rm comp},
  \label{eq:fsol}
\end{equation}
so that \(f_{\rm sol}=1\) gives purely solenoidal driving and \(f_{\rm sol}=0\) purely compressive driving. We use \(f_{\rm sol}=1\), \(0.5\), and \(0\), and label the three runs ``solenoidal'', ``mixed'', and ``compressive'' throughout. We additionally analyze a \(2048^3\) solenoidally driven realization for a convergence study.

The weight in Eq.~(\ref{eq:fsol}) acts on the amplitudes, and the two parts
are generated independently, so their cross term averages to zero and the
compressive share of the injected energy is
\begin{equation}
  \zeta = \frac{(1-f_{\rm sol})^{2}\,\sigma_{\rm comp}^{2}}
               {f_{\rm sol}^{2}\,\sigma_{\rm sol}^{2}
               +(1-f_{\rm sol})^{2}\,\sigma_{\rm comp}^{2}},
  \label{eq:zeta}
\end{equation}
where \(\sigma_{\rm sol}\) and \(\sigma_{\rm comp}\) are the rms amplitudes of the two parts as generated. For each run, we analyze three snapshots once the global multiphase statistics have reached a statistically steady state. The three snapshots are separated by one eddy turnover time of 5~Myr so that they are statistically independent. All statistics reported below are averaged over these three snapshots, and the snapshot-to-snapshot scatter is used as the uncertainty estimate. Table~\ref{tab:sim-properties} summarizes the parameters.

\begin{figure*}
\centering
\includegraphics[width=0.99\linewidth]{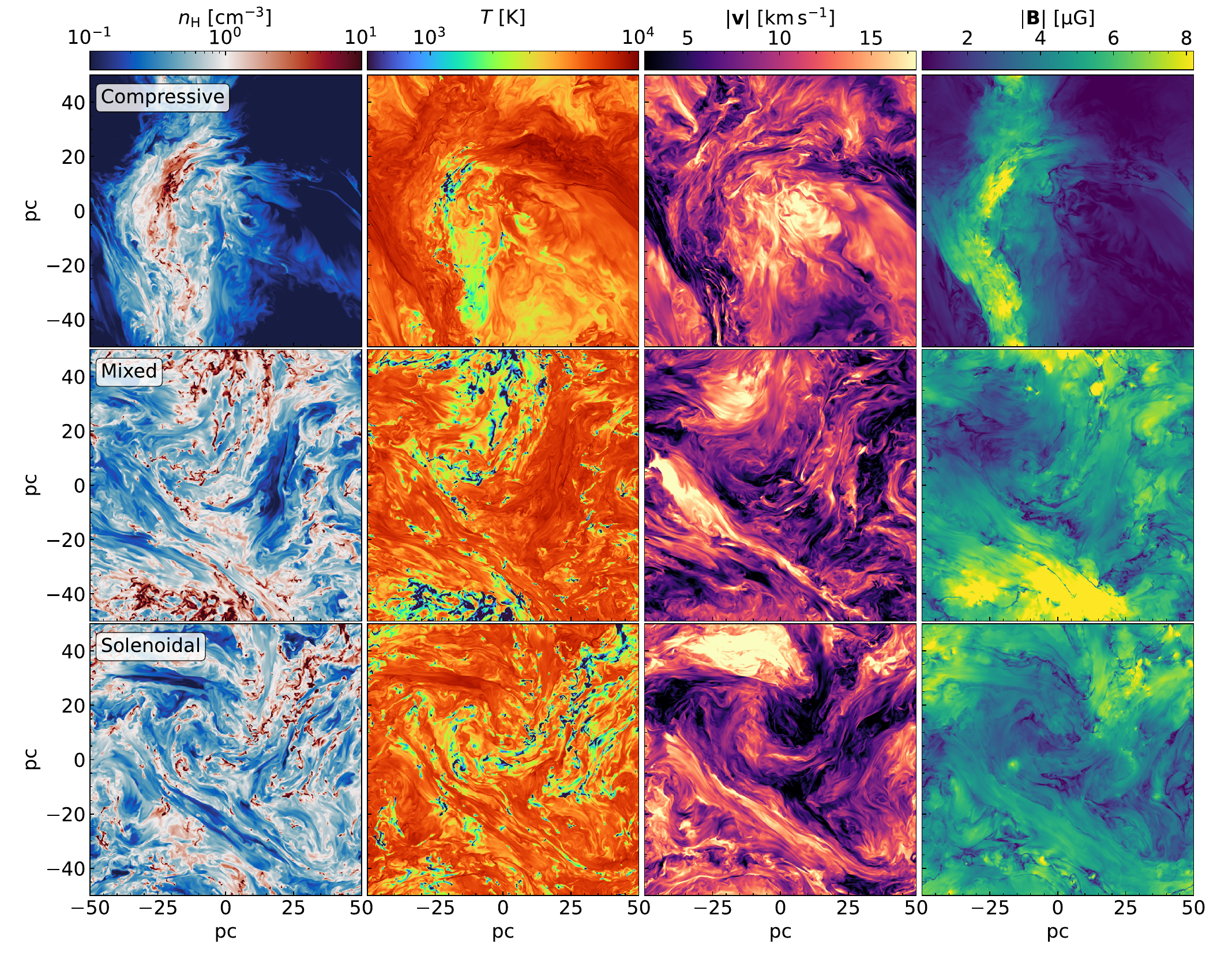}
\caption{Two-dimensional midplane slices comparing the compressively driven (top row), mixed-driven (middle row), and solenoidally driven (bottom row) simulations. From left to right: gas number density \(\nh\) (cm\(^{-3}\), log scale), temperature \(T\) (K, log scale), velocity magnitude \(|\boldsymbol{v}|\) (\(\kms\)), and magnetic field magnitude \(|\boldsymbol{B}|\) (\SI{}{\micro G}).}
\label{fig:slices}
\end{figure*}

\subsection{Increment statistics and structural conditioning}
\subsubsection{Increment definitions and phase conditioning}
\label{sec:increments}

We characterize intermittency using two-point increments separated by a lag $\boldsymbol{\ell}$. For a scalar field $s(\boldsymbol{x})$, the increment is
\begin{equation}
\delta_\ell s(\boldsymbol{x}) = s(\boldsymbol{x}+\boldsymbol{\ell}) -s(\boldsymbol{x}),
\label{eq:scalarincrement}
\end{equation}
where periodic wrapping is applied at the domain boundaries. For density we analyze both $\delta_\ell n_{\rm H}$ and the logarithmic increment
\begin{equation}
\delta_\ell\ln n_{\rm H}=\ln n_{\rm H}(\boldsymbol{x}+\boldsymbol{\ell})-\ln n_{\rm H}(\boldsymbol{x}),
\label{eq:logdensityincrement}
\end{equation}
which distinguishes absolute density contrasts from relative density variations. For a vector field $\boldsymbol{u}\in\{\boldsymbol{v},\boldsymbol{B}\}$, we use the magnitude of the full vector increment,
\begin{equation}
|\delta_\ell\boldsymbol{u}| = \left[\sum_{i\in\{x,y,z\}}(\delta_\ell u_i)^2\right]^{1/2},
\label{eq:vecmag}
\end{equation}
to calculate flatness and to select extreme vector increments. For the standardized velocity and magnetic-field PDFs, we retain the signed Cartesian component increments $\delta_\ell u_i$. Within each lag, phase, and driving sample, each component is divided by its own standard deviation, $\sigma_{\ell,i}$, and the three standardized component samples, $\delta_\ell u_i/\sigma_{\ell,i}$ with $i\in\{x,y,z\}$, are pooled into one PDF. This construction preserves the signed one-component statistics without introducing coordinate-dependent cancellations from a component sum. Unless stated otherwise, scalar statistics use $\delta_\ell n_{\rm H}$; $\delta_\ell\ln n_{\rm H}$ is used as a dynamic-range control. 

\subsubsection{Scale-dependent flatness}
\label{sec:flatness}

Our primary intermittency diagnostic is the scale-dependent increment flatness,
\begin{equation}
\Fflat(\ell)
=
\frac{S_4(\ell)}{[S_2(\ell)]^2}
=
\frac{\langle|\delta_\ell|^4\rangle}
     {\langle|\delta_\ell|^2\rangle^2}.
\label{eq:flatness}
\end{equation}
Here, $\delta_\ell$ denotes a scalar increment, $\delta_\ell n_{\rm H}$ or $\delta_\ell\ln n_{\rm H}$, for density and the magnitude of the full vector increment $|\delta_\ell\boldsymbol{u}|$ for velocity and magnetic field. An increase in $\Fflat(\ell)$ toward small scales indicates that extreme increments become increasingly important relative to the width of the distribution and is therefore a standard signature of intermittency.

For a zero-mean Gaussian scalar increment, $\Fflat_{\mathrm{G}}^{\mathrm{scalar}}=3$. For the magnitude of a three-component vector increment with independent, isotropic Gaussian components, $\Fflat_{\mathrm{G}}^{\mathrm{vector}}=5/3$. We use these values as the Gaussian references for density and for the velocity and magnetic field,
respectively. Because vector increments in magnetized turbulence need not be independent or isotropic, $5/3$ should be interpreted as a fiducial reference rather than an exact null value. Flatness above the corresponding reference indicates enhanced tails, while its variation with $\ell$ quantifies their scale dependence.

\subsubsection{Structural diagnostics of extreme increments}
\label{sec:enrichment}
The flatness quantifies how strong the extreme increments are, but not where they occur. To identify the local environment that hosts them, we pair each increment with a diagnostic evaluated at the midpoint of its separation and compare that diagnostic on the most extreme increments with its mean over the same sample.

For a given field, lag \(\ell\), and thermal phase, we sample random pairs, requiring both endpoints to lie in that phase for the WNM, UNM, and CNM subsamples (the intervening cells may belong to another phase); the Total sample is unmasked. The increment is the endpoint difference, \(|\delta_\ell n_{\rm H}|\), \(|\delta_\ell\boldsymbol{v}|\), or \(|\delta_\ell\boldsymbol{B}|\), with velocity and magnetic increments taken as the magnitude of the full vector difference, and we select the pairs in the upper 5\% of that sample. The diagnostics below are evaluated at the midpoint of each pair on the same scale \(\ell\). The enrichment is
\begin{equation}
  E(\ell) = \frac{\langle \mathcal{D}_\ell \mid |\delta_\ell| > Q_{95}\rangle}
                 {\langle \mathcal{D}_\ell \rangle},
  \label{eq:enrichment}
\end{equation}
with both averages taken over the same sample, so that changing the phase or the forcing changes both the extreme subset and its reference population. By construction \(E=1\) when the extreme increments have the same mean diagnostic as the sample as a whole, and \(E>1\) when they preferentially occupy regions of large \(\mathcal{D}_\ell\). The curves show the mean over the three separation directions and over snapshots of a given driving.

We use eight scale-matched diagnostics: three paired with density increments,
two with velocity increments, and three with magnetic-field increments. Each
diagnostic \(\mathcal{D}_\ell\) is evaluated at the midpoint on the same lag
\(\ell\), and its enrichment is computed by conditioning on extreme
increments within the chosen phase sample and normalizing by the diagnostic's
own mean over that sample.

\textbf{For density increments,}
\begin{equation}
\begin{aligned}
  C_{\ell,+} &= \max\left(-\boldsymbol{\nabla}_\ell\cdot\boldsymbol{v},\,0\right),
  \qquad
  \mathcal{D}_{\nabla T} = |\boldsymbol{\nabla}_\ell T|,\\
  S_\ell^{2} &= \sum_{ij} S_{ij}S_{ij},
  \qquad
  S_{ij} = \tfrac{1}{2}\left(\partial_i v_j + \partial_j v_i\right)-\frac{1}{3}\delta_{ij}\boldsymbol{\nabla}_\ell\cdot\boldsymbol{v},
\end{aligned}
  \label{eq:diag_density}
\end{equation}
which isolate convergent flow, thermal interfaces, and shear-driven
deformation, respectively. Comparing the enrichment of \(C_{\ell,+}\) with
that of \(S_\ell^{2}\)---the same traceless strain used for velocity
increments, here paired with density instead---tests whether extreme density
fluctuations are hosted by compression or by shear, without double-counting
the compressive contribution already captured by \(C_{\ell,+}\).

\textbf{For velocity increments,}
\begin{equation}
      \Omega_\ell^{2} = |\boldsymbol{\nabla}_\ell\times\boldsymbol{v}|^{2},
  \qquad
  S_\ell^{2} = \sum_{ij} S_{ij}S_{ij},
  \label{eq:diag_velocity}
\end{equation}
the squared vorticity and the squared traceless strain, which isolate
rotation and shear-driven deformation and exclude the isotropic compression
already captured by \(C_{\ell,+}\). Comparing their enrichments tests whether
extreme velocity increments are hosted by rotation- or by shear-dominated
structures.

\textbf{For magnetic-field increments,}
\begin{equation}
\begin{aligned}
   J_\ell^{2} = |\boldsymbol{\nabla}_\ell\times\boldsymbol{B}|^{2}&,
  \qquad
  \tau_\ell^{2} = \left|(\boldsymbol{B}\cdot\boldsymbol{\nabla}_\ell)\boldsymbol{B}\right|^{2},\\\\
 & C_{B,\ell} = |\boldsymbol{B}|\, C_{\ell,+},
\end{aligned}
  \label{eq:diag_magnetic}
\end{equation}
the squared scale-matched magnetic-curl proxy, the squared scale-matched magnetic-tension proxy, and the magnetic compression proxy, and the amplitude of the compressive source term
in the induction equation. The three probes distinct origins of magnetic fluctuations---current structure, tension, and compression---so a large enrichment in one does not imply a large enrichment in the others.

Because \(E\to1\) can indicate either that the extreme increments have no structural preference or that the diagnostic itself varies little across the sample, and because the eight diagnostics differ in dimension and functional form, each is normalized by its own mean: the value of \(E\) measures how strongly a diagnostic is concentrated on the extreme increments, not an amplitude shared across diagnostics.

\section{Results}\label{sec:results}

\subsection{Forcing-dependent phase structure and budget}
\label{sec:global}
Fig.~\ref{fig:slices} shows midplane slices of the hydrogen number density \(n_{\rm H}\), temperature, velocity magnitude, and magnetic-field strength for the three runs. Compressive driving produces sharper density boundaries and more extended, coherent structures, with the cold gas gathered into a single large-scale band and the remainder of the domain left comparatively rarefied.
The solenoidal run instead distributes a larger fraction of the cold gas into small-scale structures spread through the volume. The mixed run resembles the solenoidal case more closely than the compressive one, consistent with its compressive energy share being well below one half (Eq.~\ref{eq:zeta}). The same ordering appears in the magnetic field: the compressive run concentrates the strongest field within the dense band, whereas the mixed and solenoidal runs show a more uniformly distributed field.

These morphological differences are reflected in the second-order structure functions (Fig.~\ref{fig:s2} in the Appendix), but not primarily through their slopes. For \(\ell\lesssim1\)~pc all three runs follow comparable scalings, bracketed by \(\ell^{2/3}\) and \(\ell^{1}\). The runs differ instead in amplitude and in the scale at which the density structure functions saturate. The total-gas density amplitude is about five times larger under compressive driving, and the saturation, which sets in once the separation exceeds the characteristic size of the dominant density structures, occurs at \(\ell\simeq3\)~pc in the solenoidal and mixed runs.

\begin{figure*}
\centering
\includegraphics[width=0.99\linewidth]{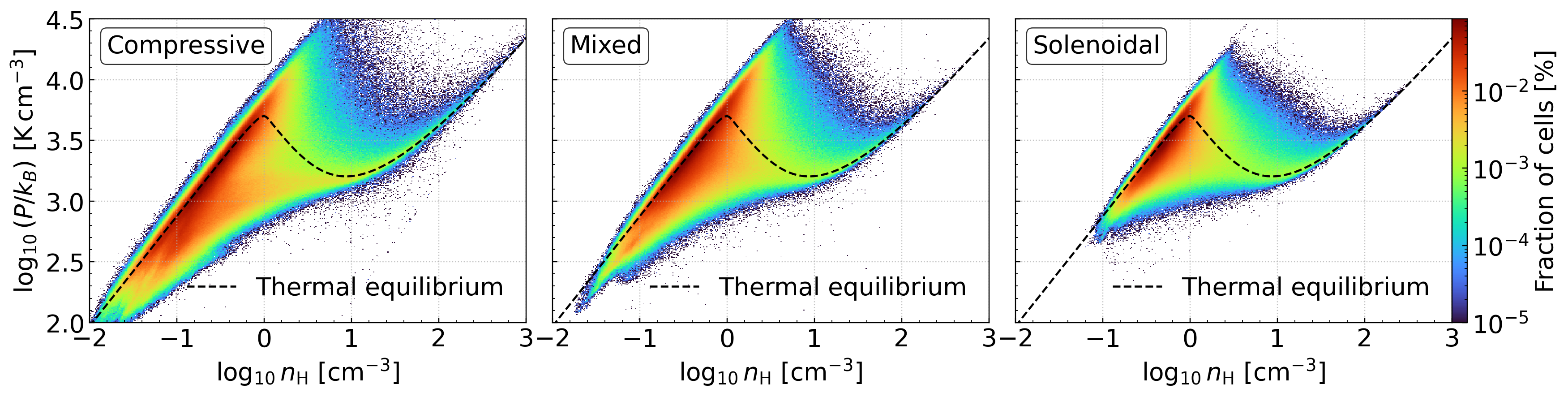}
\caption{Two-dimensional phase diagrams of gas pressure \(P/k_B\) (K\,cm\(^{-3}\)) versus number density \(\nh\) (cm\(^{-3}\)) for the compressively driven (left), mixed driven (middle), and solenoidally driven (right) simulations. The color scale gives the fraction of computational cells occupying each bin. The black dashed curve is the thermal equilibrium locus of the adopted heating and cooling functions, obtained by solving \(\Gamma=\nh\Lambda(T)\) at each \(\nh\). }
\label{fig:phasediag}
\end{figure*}

\begin{figure*}
\centering
\includegraphics[width=0.99\linewidth]{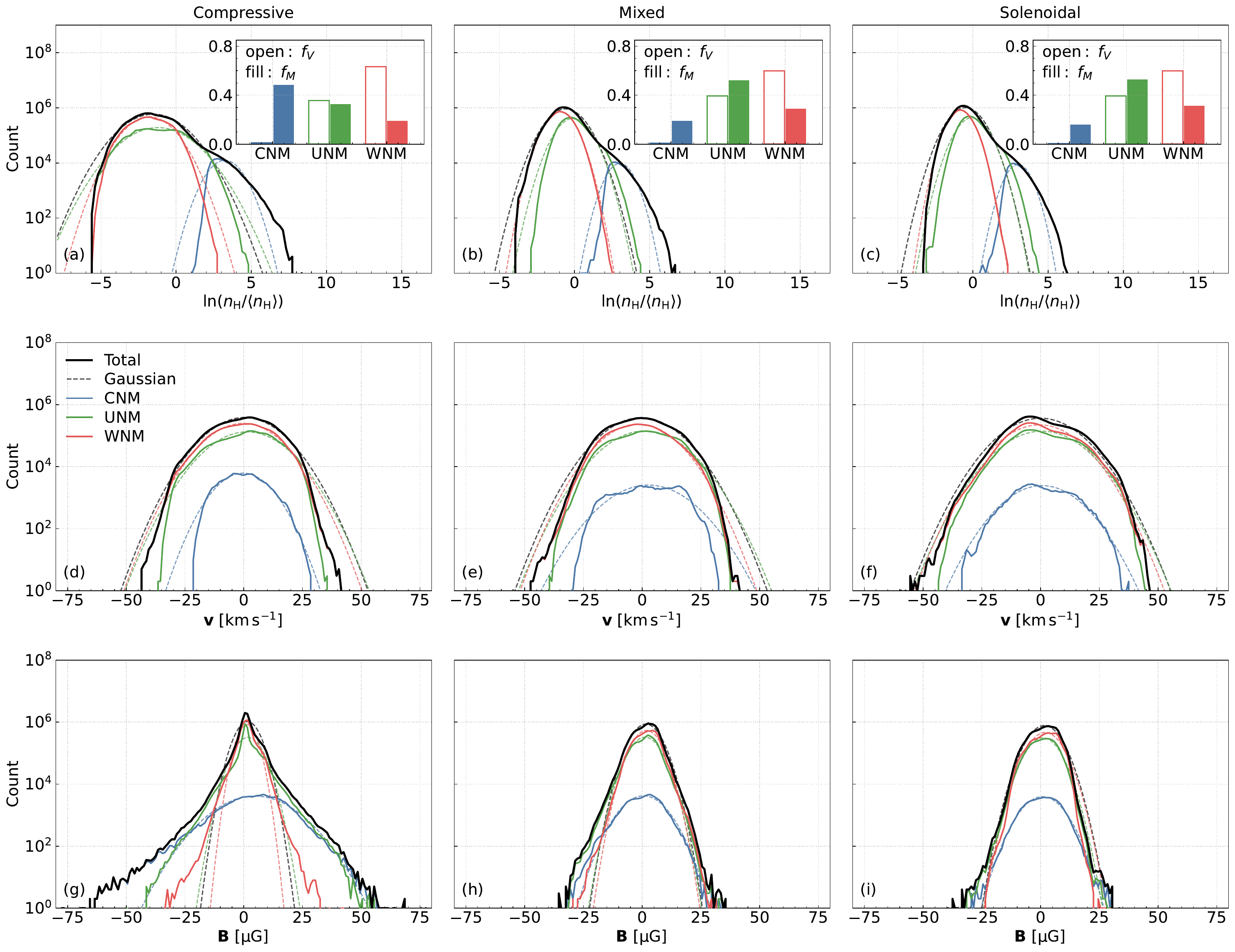}
\caption{One-point distributions of density, velocity, and magnetic field for the compressively driven (left column; panels a, d, g), mixed-driven (middle column; panels b, e, h), and solenoidally driven (right column; panels c, f, i) simulations, decomposed by thermal phase (CNM, blue; UNM, green; WNM, red) and compared with the total (black). (a, b)
Distribution of the logarithmic density contrast \(\ln(\nh/\langle \nh\rangle)\); inset bars give the volume filling fraction \(f_V\) (open) and mass fraction \(f_M\) (filled) of each phase. (c, d) Line-of-sight velocity, with dashed curves showing Gaussians of the same variance. (e, f) The corresponding magnetic-field component distributions.}
\label{fig:onepoint}
\end{figure*}

\subsection{Thermodynamic phase structure}
The ISM is modeled as a multiphase medium comprising a warm, diffuse component with temperatures $T > 5000$~K (WNM); an intermediate-density, thermally unstable component with $200 < T < 5000$~K (UNM); and a cold, dense component with $T < 200$~K (CNM). The phase definition is motivated by the thermal-equilibrium curve in the \(P\)--\(n\) phase diagram: the stable WNM and CNM branches have
\(dP/dn>0\), whereas the thermally unstable branch has \(dP/dn<0\), as shown in Fig.~\ref{fig:phasediag}. The two turning points separating these branches occur at approximately \(T\simeq 5000\)~K and \(T\simeq 200\)~K, which set the temperature boundaries adopted here \citep{1977ApJ...218..148M,2003ApJ...587..278W,2011piim.book.....D}. 

All runs populate the expected cold, unstable, and warm regimes, but the compressive run has the largest departures from thermal equilibrium, including both low-pressure diffuse gas and a high-pressure tail. It also reaches \(n_{\rm H}\simeq10^{-2}~{\rm cm^{-3}}\), about an order of magnitude below the other runs. The mixed and solenoidal runs remain closer to the equilibrium locus. These differences show that a temperature-selected phase samples different dynamical conditions under different forcing, motivating the phase-conditioned comparison below.

\subsection{One-point probability distributions}
The one-point Probability Distribution Functions (PDFs) in Fig.~\ref{fig:onepoint} show pronounced non-Gaussian fluctuations and how their occurrence depends on both forcing and thermal phase. The total density PDF is considerably broader in the compressive run, for which \(\sigma_{\ln\nh}=1.3\), than in the solenoidal run, for which \(\sigma_{\ln\nh}=0.8\); the mixed run, at \(0.9\), lies close to the solenoidal case. The difference by a factor of \(1.6\) is consistent in sign and magnitude with the forcing dependence found in isothermal simulations
\citep{2008ApJ...688L..79F,2010A&A...512A..81F}. The broadening is not symmetric: it occurs mainly on the rarefied side, where the compressive distribution peaks near \(\ln(\nh/\langle\nh\rangle)\simeq-2\) and extends to \(\simeq-7\), whereas the other two runs peak near \(-1\) and are truncated near \(-4\). The dense side is extended only modestly by comparison. The same asymmetry appears in Fig.~\ref{fig:phasediag}, where only the compressive run reaches \(\nh\simeq10^{-2}~{\rm cm^{-3}}\), indicating that the broader
density distribution reflects rarefaction as much as compression.

The phase-conditioned PDFs show that compressive driving increases both the separation between the thermal phases and the density variation within each phase. The characteristic densities of the WNM, UNM, and CNM are \(0.3\), \(1.0\), and \(35.2~{\rm cm^{-3}}\) under compressive driving, against \(0.5\), \(1.3\), and \(25.2~{\rm cm^{-3}}\) in the solenoidal run, so the two diffuse phases are displaced to lower densities and the CNM to higher ones. The corresponding logarithmic dispersions are \(1.0\), \(1.3\), and \(0.8\) under compressive driving, larger than the values \(0.6\), \(0.7\), and \(0.6\) obtained under solenoidal driving. Compressive driving therefore broadens the density distribution within every phase while pulling the phases further apart.

\begin{figure*}
\centering
\includegraphics[width=0.99\linewidth]{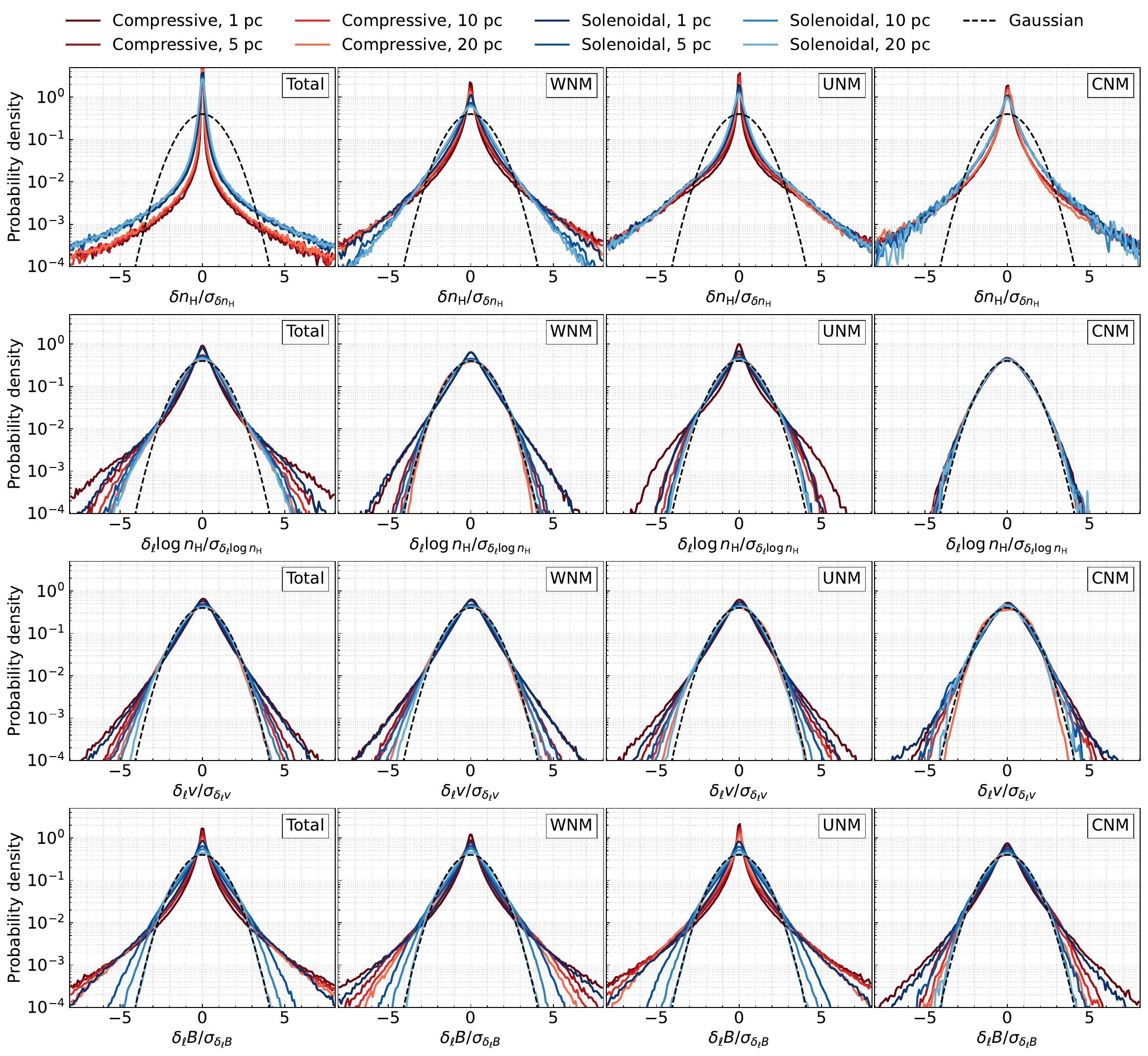}
\caption{Standardized increment PDFs at four lags (\(\ell=1,5,10,20\)~pc) for density (1st row), logarithmic density (2nd row), velocity (3rd row), and magnetic field (4th row), evaluated for the total gas and for the WNM, UNM, and CNM phases (left to right). Dark to light shading indicates increasing lag; red curves are the compressive run and blue curves the solenoidal run. Each distribution is normalized by its own standard deviation, so the comparison is of shape only, and the dashed black curve is a unit Gaussian. Density and logarithmic density use scalar increments. For velocity and magnetic field, the PDF pools the signed Cartesian-component increments after each component has been normalized by its own standard deviation.}
\label{fig:pdfs}
\end{figure*}

The volume and mass fractions in the insets separate the roles of the phases. The CNM occupies \(1\%\) of the volume in all three runs, but its mass fraction depends strongly on the forcing: \(f_M=0.43\) under compressive driving against \(0.22\) and \(0.17\) in the mixed and solenoidal runs. The WNM fills \(\simeq60\%\) of the volume throughout while carrying \(0.19\)--\(0.31\) of the mass, and the UNM occupies most of the remainder, carrying \(0.38\) of the mass under compressive driving and \(0.51\)--\(0.53\)
in the other two runs. The rare high-density structures of the CNM are thus volumetrically negligible while carrying a mass fraction that the forcing changes by a factor of \(2.5\), largely at the expense of the UNM. The broad total density PDF therefore reflects both fluctuations within individual phases and the superposition of thermodynamically distinct components.

The velocity PDFs show only a weak dependence on either forcing or phase. The volume-mean velocity magnitude is in fact slightly smaller under compressive driving, \(9.1~\kms\) against \(10.0~\kms\) in the solenoidal run, and the
phase-to-phase variation within a given run does not exceed \(\simeq10\%\) (Table~\ref{tab:sim-properties}). The strongly supersonic CNM, with \(\mathcal{M}_s=7.4\)--\(8.5\), therefore arises entirely from its low sound speed rather than from large velocity amplitudes; its velocity magnitude is comparable to that of the warm gas in every run.

The magnetic-field PDFs behave differently again, and in a way that the global mean conceals. The volume-mean field strength is \emph{lower} under compressive driving, \(3.1~\SI{}{\micro G}\) against \(4.7~\SI{}{\micro G}\) in the solenoidal run, yet this deficit is confined to the diffuse gas: the WNM and UNM means fall from \(4.7~\SI{}{\micro G}\) to \(3.0\) and \(3.1~\SI{}{\micro G}\), while the CNM mean rises from \(6.0\) to \(9.5~\SI{}{\micro G}\). Compressive driving thus amplifies the global field less efficiently than the mixed and solenoidal driving, but it redistributes magnetic energy into the \(1\%\) of the volume occupied by cold gas, leaving the remaining \(99\%\) more weakly magnetized. The shape of the distribution changes accordingly: the compressive run develops non-Gaussian wings reaching \(|\boldsymbol{B}|\gtrsim50~\SI{}{\micro G}\), well beyond the Gaussian of the same variance, whereas the mixed and solenoidal distributions remain compact. The corresponding plasma \(\beta\) in the CNM falls to \(0.1\), against \(0.2\) in the other two runs, so the cold gas under compressive driving is the most strongly magnetized relative to its thermal pressure. This concentration of magnetic energy into a small volume fraction is a one-point signature of the intermittent magnetic structure examined below.

\subsection{Scale-dependent increment distributions}
\label{sec:pdfs}
Fig.~\ref{fig:pdfs} shows the standardized increment PDFs at four lags, \(\ell=1\), \(5\), \(10\), and \(20\)~pc. The density and logarithmic-density rows use signed scalar increments. The velocity and magnetic-field rows pool signed, separately standardized Cartesian-component increments, so each vector PDF can be compared directly with a unit Gaussian without component cancellation. The figure therefore compares distribution shapes rather than absolute fluctuation amplitudes. All three fields are non-Gaussian at every lag and in every phase, with sharp central peaks and tails extending well beyond the Gaussian reference. The lag dependence, however, is not the same for all fields: the velocity and magnetic-field tails broaden toward small \(\ell\), the standard signature of a turbulent cascade, whereas the density tails broaden toward \emph{large} \(\ell\).

The density increments are the most strongly non-Gaussian. Their tails are widest at \(\ell=20\)~pc and narrowest at \(\ell=1\)~pc, opposite to the behavior of the velocity and magnetic fields. This reversal follows from the structure of the medium rather than from the cascade: at small lags both points of an increment usually sample the same coherent structure, whereas at larger lags an increasing fraction of pairs straddles a phase boundary and registers the full cold-to-warm density contrast. Consistent with this
interpretation, the total-gas PDF is the most sharply peaked and the most extended of the four panels, since it alone retains these cross-phase pairs. Among the phase-conditioned distributions, the WNM is closest to Gaussian, the UNM is intermediate, and the CNM has both the sharpest peak and the heaviest tails. The ordering between the compressive and solenoidal curves varies with phase and lag, so compressive driving does not produce uniformly heavier standardized density tails; the forcing imprint depends on which thermal
component is sampled.

The velocity increments are closer to Gaussian than their density counterparts, particularly at \(\ell=10\)--\(20\)~pc, where the distributions nearly coincide with the reference. Toward smaller lags, the tails broaden progressively toward smaller lags, consistent with scale-dependent intermittency. The scale dependence is present under both forcings, but the compressive run develops markedly stronger tails at \(\ell=1\)~pc in the total gas, WNM, and UNM. In the CNM, the two forcings overlap closely at all lags, and the CNM velocity
distributions are the least extended of the four panels.

The magnetic-field increments show the clearest scale dependence of the three. At \(\ell=1\)~pc, the PDFs combine a sharp central peak with broad wings, and they narrow steadily toward the Gaussian reference as the lag increases. The small-scale wings are more prominent under compressive driving in the total gas, WNM, and UNM, whereas in the CNM the forcing dependence is weaker and less systematic. This progression is consistent with magnetic fluctuations becoming concentrated into increasingly rare and intense structures toward small scales. Taken together, the three fields show that the non-Gaussianity of the multiphase ISM depends on the scale, the thermal phase, and the forcing, and that these dependences are not the same for the density, velocity, and magnetic fields.

\begin{figure*}
\centering
\includegraphics[width=0.99\linewidth]{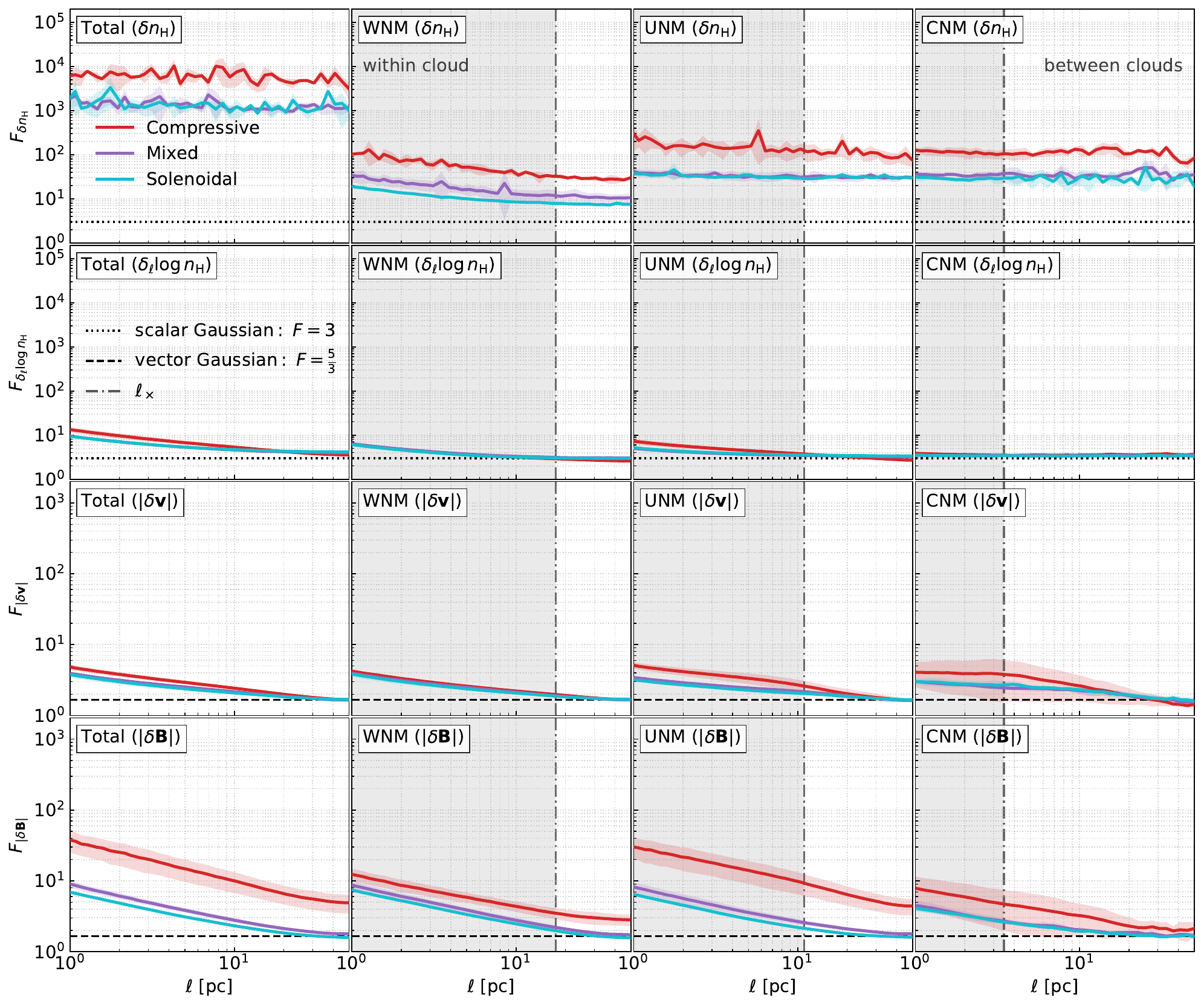}
\caption{Increment flatness \(\Fflat=S_4/S_2^2\) as a function of spatial lag \(\ell\) for density (1st row), logarithmic density (2nd row), velocity (3rd row), and magnetic field (4th row), evaluated for the total gas and for the WNM, UNM, and CNM phases (left to right). Red curves are the compressive run, cyan the solenoidal run; shaded envelopes span the three separation directions. Horizontal lines mark the Gaussian references, \(\Fflat_{\rm G}=3\) for the scalar density (dotted) and \(5/3\) for the vector magnitudes (dashed). Velocity and magnetic field use the full magnitude of the vector increment. The vertical dash-dotted line in each phase-conditioned panel marks the coherence scale \(\ell_\times\) of that phase; shading separates the within-cloud (\(\ell<\ell_\times\)) and between-cloud (\(\ell>\ell_\times\)) regimes for the WNM/UNM and CNM, respectively.}
\label{fig:flatness}
\end{figure*}

\begin{figure*}
\centering
\includegraphics[width=0.99\linewidth]{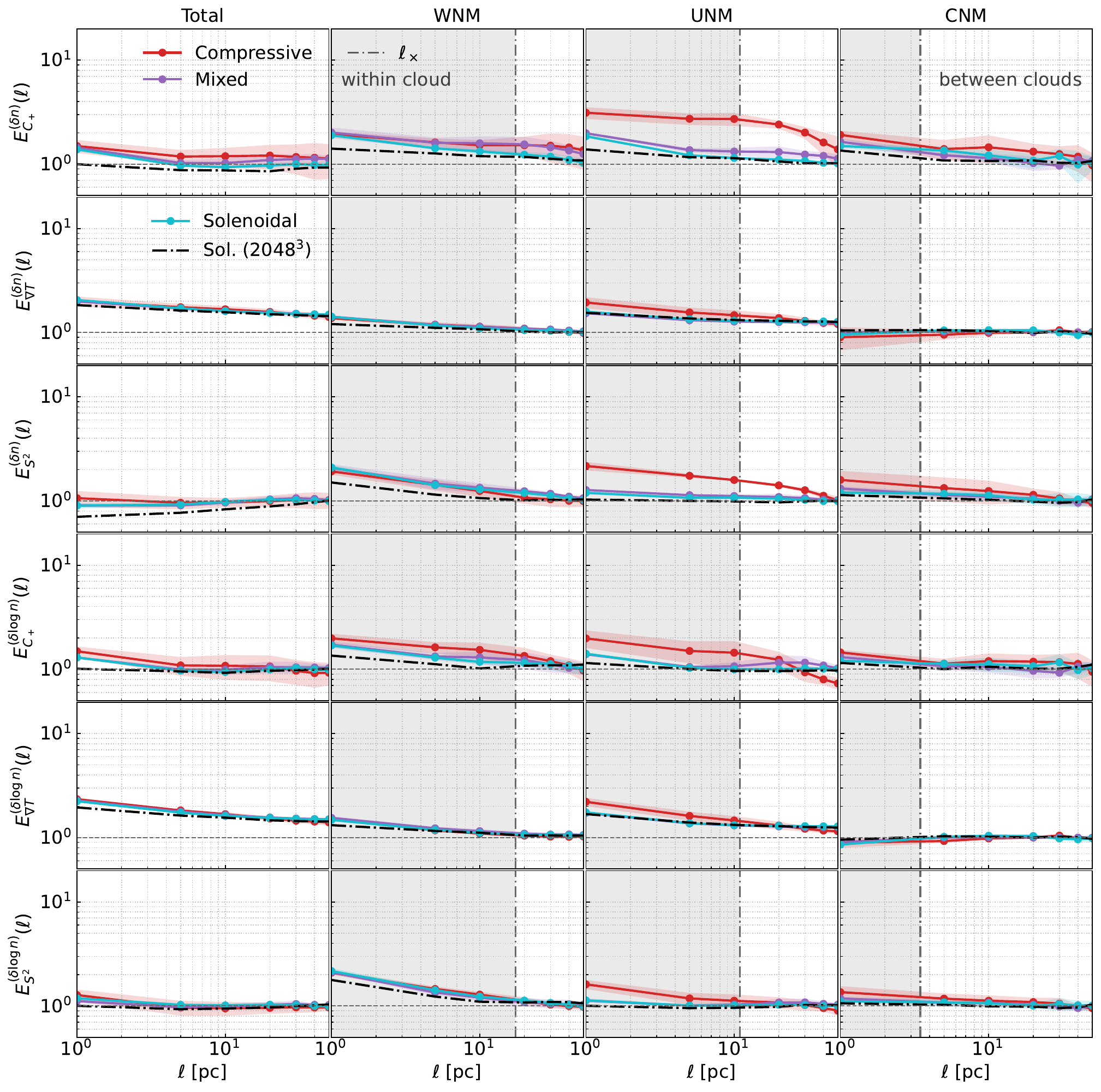}
\caption{Conditional enrichment for the extreme density (1st - 3rd rows) and logarithmic density (4th - 6th row) increments, \(E_{C_+}\), \(E_{\nabla T}\) (middle), and \(E_{S^2}\), as a function of lag \(\ell\), for the total gas and for the WNM, UNM, and CNM phases (left to right). Red, purple, and cyan mark the compressive, mixed, and \(1024^3\) solenoidal runs; shaded bands give the standard deviation over the three snapshots of each driving. Black dash-dotted curves show the corresponding \(2048^3\) solenoidal result. The horizontal dashed line marks \(E=1\), the value expected if the extreme increments have no structural preference. The vertical dash-dotted line in each phase-conditioned panel marks the coherence scale \(\ell_\times\) of that phase; shading separates the within-cloud (\(\ell<\ell_\times\)) and between-cloud (\(\ell>\ell_\times\)) regimes for the WNM/UNM and CNM, respectively.}
\label{fig:e_density}
\end{figure*}

\subsection{Scale-dependent intermittency}
\label{sec:flatnessresults}
Fig.~\ref{fig:flatness} shows the increment flatness as a function of lag for the total gas and the three thermal phases. The horizontal lines mark the Gaussian references, \(\Fflat_{\rm G}=3\) for density and \(\Fflat_{\rm G}=5/3\) for the full velocity- and magnetic-field-increment magnitudes, and the vertical line in each phase-conditioned panel marks the coherence scale \(\ell_\times\) of that phase. We define \(\ell_\times\) as the lag at which the magnetic second-order structure function of that phase reaches \(80\%\) of its saturation value (Fig.~\ref{fig:s2}); the magnetic field is used because it is the only one of the three fields whose structure function saturates cleanly, and at a comparable lag across drivings, in every phase. Beyond \(\ell_\times\), same-phase pairs increasingly sample different structures of that phase (``between clouds'') rather than fluctuations within one (``within cloud''). For all three fields, the flatness generally increases toward smaller scales and remains above its Gaussian reference throughout most of the resolved range, showing the same scale-dependent non-Gaussianity identified in Fig.~\ref{fig:pdfs}. The mixed run tracks the solenoidal one closely in every panel.

The density increments are by far the most non-Gaussian, with \(\Fflat_{n_{\rm H}}\) reaching \(10^{3}\)--\(10^{4}\) in the total gas over \(1\lesssim\ell\lesssim10\)~pc. Conditioning both endpoints on the same thermal phase reduces the flatness by roughly an order of magnitude, showing that contrasts across phase boundaries strongly contribute to the extreme increments in the total density field; even so, every phase-conditioned distribution remains non-Gaussian. The UNM and CNM reach comparable, and the largest, phase-conditioned values, while the WNM has the lowest flatness under both forcings. The \(\delta_\ell\ln n_{\rm H}\) control suppresses the disproportionate fourth-moment weight of the densest gas; it should therefore be used to assess scale dependence and phase ordering rather than the absolute value of \(\Fflat_{n_{\rm H}}\). The coherence scale is smallest in the CNM and larger in the WNM and UNM; beyond it, the CNM forcing contrast approaches that of the total-gas curve, consistent with the two samples increasingly probing the same between-cloud medium.

The velocity increments show weaker but still clear intermittency. Their flatness decreases smoothly with lag and approaches the vector Gaussian
reference at large scales. Compressive driving produces systematically larger flatness in the total gas, WNM, and UNM throughout the resolved range.
In the CNM, the two forcings remain close at small \(\ell\) but separate again beyond \(\ell_\times\), where the between-cloud statistics retain an elevated flatness out to the largest lags probed. Compared with density, the phase-to-phase variation is modest, while the forcing contrast is clearest in the warm and unstable gas.

The magnetic-field increments show the clearest forcing dependence of the three fields. Under compressive driving, the total-gas and UNM flatness
remain of order tens over much of the resolved range, well above the solenoidal and mixed curves; the WNM and CNM show the same enhancement but
with a smaller contrast. The strongest phase-conditioned magnetic intermittency occurs in the UNM rather than the CNM, so the phase with the
broadest one-point magnetic fluctuations (Section~\ref{sec:pdfs}) is not the one with the strongest scale-dependent intermittency. In every phase, the flatness declines toward \(5/3\) at large lags while the forcing contrast grows toward small scales, and in the CNM this decline is interrupted near \(\ell_\times\), consistent with the same transition identified in the density and velocity panels.

We restrict quantitative comparisons to \(1~{\rm pc}\leq\ell\leq50~{\rm pc}\): the region \(\ell<1\)~pc is affected by numerical dissipation, and lags approaching or exceeding \(\ell_\times\) increasingly probe the medium between structures rather than a single one. Within this range,
Fig.~\ref{fig:flatness} shows that forcing and thermal phase both leave
scale-dependent imprints on intermittency, with a consistent
phase-dependent transition scale recurring across all three fields.

\begin{figure*}
\centering
\includegraphics[width=0.99\linewidth]{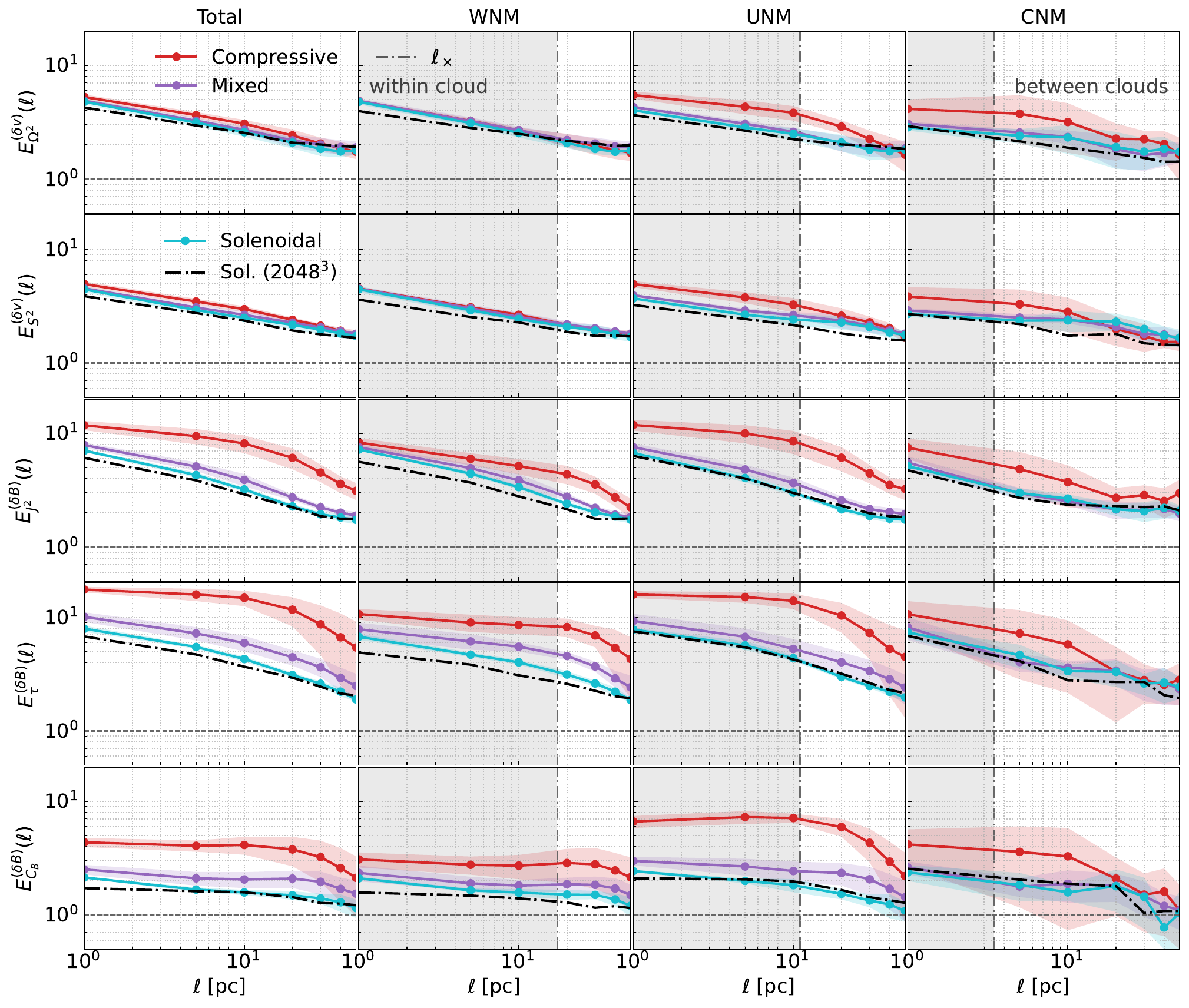}
\caption{Conditional enrichment for the extreme increments of velocity (\(E_{\Omega^2}\)
and \(E_{S^2}\), top two rows) and magnetic field (\(E_{J^2}\), \(E_\tau\), and \(E_{C_B}\), bottom three rows), as a function of lag \(\ell\), for the total gas and for the WNM, UNM, and CNM phases (left to right). The black dash-dotted curves are the \(2048^3\) solenoidal results; all other colors, shading, and reference lines are as in Fig.~\ref{fig:e_density}.}
\label{fig:e_vB}
\end{figure*}

\subsection{Structural hosts of extreme increments}
Figs.~\ref{fig:e_density} and~\ref{fig:e_vB} show the conditional enrichment \(E(\ell)\) of the eight diagnostics for the extreme increments of density, velocity, and magnetic field. Because \(C_+\) and \(\nabla T\) are first order in the derivatives while the remaining four diagnostics are second order, absolute values of \(E\) are compared only within each order, as discussed in Section~\ref{sec:enrichment}.

The density diagnostics show only modest enrichment. \(E_{C_+}\) exceeds unity by a factor of \(2\)--\(3\) in the UNM under compressive driving but is close to unity elsewhere, including in the CNM at all lags; \(E_{\nabla T}\) behaves similarly, peaking at \(\simeq2\) in the UNM and remaining near unity in the CNM throughout. \(E_{S^2}\), the traceless-shear control, tracks \(E_{C_+}\) closely in shape but not in strength, staying within a factor of \(2\) of unity even where \(E_{C_+}\) is largest. The conventional picture of compressible turbulence attributes extreme density fluctuations to shocks and converging flows \citep[e.g.,][]{2008ApJ...688L..79F}; our result matches this expectation in the UNM, where \(E_{C_+}\) is the most enhanced diagnostic among the three, but not in the CNM, where none of the three diagnostics departs from unity at any scale probed. The coldest, densest gas is thus the phase in which the extreme density increments show the weakest structural preference of all, contrary to the expectation that the least compressible phase should be the most shock-dominated. The \(2048^3\) solenoidal curves reproduce the near-unity CNM enrichment and the weak density-diagnostic ordering of the \(1024^3\) run. This agreement supports the conclusion at the resolved lags, while leaving open sub-pc structure and the forcing dependence of resolution effects.

The velocity and magnetic-field diagnostics in Fig.~\ref{fig:e_vB} show a markedly different picture, and one that departs from textbook expectations. \(E_{\Omega^2}\) and \(E_{S^2}\) for the velocity increments are comparable to each other at every lag and in every phase, both reaching \(\simeq4\)--\(6\) at \(\ell=1\)~pc and declining toward unity at large lags. The classical picture of incompressible intermittency attributes it to vortex filaments, thin tube-like regions of concentrated vorticity in which \(\Omega^2\) should dominate over the strain \(S^2\) \citep[e.g.,][]{1993PhRvE..48...29B,1994PhRvL..72..336S}; our result does not match this expectation, since the extreme velocity increments show no preference for rotation over shear. This does not support a purely vorticity-dominated filament picture. Instead, rotational and strain diagnostics are comparably enriched, indicating that extreme velocity increments involve both rotation and deformation rather than a single dominant kinematic structure. At \(2048^3\), the solenoidal curves closely track the \(1024^3\) density, kinematic, and magnetic diagnostics. This supports the field-to-field structural contrast at resolved lags.

The magnetic diagnostics are enriched far more strongly than either velocity diagnostic: \(E_{J^2}\) and \(E_\tau\) both exceed \(10\) at \(\ell=1\)~pc under compressive driving, roughly twice the peak enrichment of \(E_{\Omega^2}\) or \(E_{S^2}\), and the two track each other closely in both magnitude and forcing dependence at every lag and phase. Two distinct pictures are available for magnetic intermittency: current sheets, where \(J^2\) is large and the field reverses across a thin layer, and folded or kinked field lines, where \(\tau\) is large from curvature and stretching without necessarily concentrating current. Our result matches neither in isolation---\(E_{J^2}\) and \(E_\tau\) are not merely both enriched, but very similar in strength and behavior throughout --- indicating that extreme magnetic increments preferentially occur in regions where magnetic-curl and magnetic-tension diagnostics are simultaneously enhanced. This behavior is consistent with magnetic structures that combine strong field gradients and field-line bending, although the present diagnostics do not by themselves establish a field reversal or a specific folded-field geometry.

A similar coincidence of curvature and current has been invoked for small-scale magnetic-field reversals in large-amplitude MHD turbulence \citep{2023MNRAS.525.4985K}. Our signal is distinct in extent: the enrichment declines from its peak at \(\ell=1\)~pc but remains above unity across the full resolved range to \(\ell\simeq50\)~pc, more than an order of magnitude beyond our effective dissipation scale. The \(2048^3\) solenoidal result reproduces the joint scale dependence of \(E_{J^2}\) and \(E_\tau\), strengthening the conclusion that extreme magnetic increments preferentially occupy curved, current-carrying structures at the resolved scales. This selectivity is also the most strongly forcing-dependent among the diagnostics considered. \(E_{J^2}\) and \(E_\tau\) are enhanced under compressive driving by a factor of \(2\)--\(3\) over the mixed and solenoidal runs in the total gas, WNM, and UNM, whereas the density and velocity diagnostics show little or no such contrast.



\section{Discussion}
\label{sec:discussion}

\subsection{Density intermittency: phase transitions, chemistry, and cloud evolution}
The total-gas density PDF combines warm, unstable, and cold populations and therefore need not obey the single-component, approximately lognormal closure often used for isothermal turbulence \citep{2008ApJ...688L..79F,2010A&A...512A..81F}. Phase conditioning and the \(\delta_\ell\ln n_{\rm H}\) control nevertheless leave substantial non-Gaussianity, showing that separated phase peaks are not its sole origin. Extreme density increments preferentially coincide with convergence and thermal gradients in the UNM, especially under compressive driving, and thus identify candidate interfaces where compression, cooling, and phase conversion are coupled. Their weak instantaneous enrichment in the CNM may instead reflect the persistence of cold structures after formation.

This interpretation can be tested with matched H~I 21-cm emission and absorption, using projected column-density increments together with spin-temperature or CNM-fraction estimates. CH\(^{+}\) absorption offers a complementary, chemically selective probe because multiphase calculations associate it with turbulent CNM--WNM exchange and thermal instability \citep{2023A&A...669A..74G}. A spatial or velocity-resolved excess of CH\(^{+}\) near extreme H~I increments would support the interface interpretation. It would not uniquely identify intermittent dissipation, however, and neither observable directly recovers the three-dimensional \(\delta_\ell n_{\rm H}\).

\subsection{Velocity intermittency: shear, mixing, and energy transfer}
Extreme velocity increments show comparable vorticity and traceless-strain enrichment, so their hosts contain both rotation and shear rather than only vortex filaments. This agrees qualitatively with coherent velocity-shear structures inferred in diffuse molecular gas \citep{2008A&A...481..367H,2009A&A...500L..29H}. Observationally, the direct analogue is the tail of velocity-centroid increments: extreme pairs should preferentially overlap large centroid-gradient amplitudes and broadened or multi-component profiles. Projection, opacity, thermal broadening, and line-of-sight blending can generate or erase such associations, so synthetic position--position--velocity (PPV) cubes are required before comparison. The diagnostic locates candidate mixing and energy-transfer sites but does not measure irreversible heating; that requires filtered energy fluxes, pressure--dilatation, and Lorentz work.

\subsection{Magnetic intermittency: field topology, energy conversion, and transport}
Magnetic increments are more strongly localized than velocity increments, and their comparable magnetic-curl and magnetic-tension enrichment shows that strong gradients and field-line distortion coexist. Ideal-MHD diagnostics do not, however, establish a reversal, folded topology, reconnection, or irreversible heating. The association remains relevant to charged-particle transport, although a measured curvature scale cannot be assigned to a particle energy without comparison to the gyroradius \citep{2023MNRAS.525.4985K,2024MNRAS.528.4245B,2025ApJ...990L..18K}.

The velocity gradient technique (VGT) provides a specific observational test. Rotated velocity-centroid or channel-map gradients statistically trace the plane-of-sky magnetic orientation in MHD turbulence \citep{2018ApJ...853...96L,2018MNRAS.480.1333H,2023MNRAS.524.2379H}, and applications recover cloud-scale field morphology consistent with dust polarimetry \citep{2019NatAs...3..776H,2022MNRAS.511..829H,2022MNRAS.510.4952L}. If the simulated association survives projection, sightlines intersecting extreme magnetic-increment structures should show unusually rapid spatial changes or enhanced dispersion in the VGT-inferred orientation, potentially accompanied by reduced local agreement with polarized-dust angles. 


\section{Conclusions}
\label{sec:conclusions}

We analyzed three matched $1024^3$ multiphase ideal-MHD simulations with compressive, mixed, and solenoidal forcing, together with a $2048^3$ solenoidal realization used to test numerical convergence. We quantified the scale-dependent intermittency of density, velocity, and magnetic-field fluctuations and identified the physical environments associated with their most extreme increments. Our main conclusions are:

\begin{enumerate}

\item All three fields exhibit non-Gaussian, scale-dependent increments across the resolved range. Density shows the largest flatness, while magnetic-field intermittency has the clearest dependence on turbulent forcing and becomes increasingly pronounced toward small scales.

\item Compressive driving substantially broadens the density distribution, increases the CNM mass fraction, and redistributes both density and magnetic power preferentially into cold gas. The forcing therefore changes not only the strength of the fluctuations but also how they are partitioned among thermal phases.

\item Thermal-phase conditioning reduces the extreme density contrast present in the total medium, but substantial non-Gaussianity remains within each phase. The observed intermittency is therefore not simply a consequence of superposing distinct WNM, UNM, and CNM populations.

\item Extreme density increments preferentially occur in regions of converging flow and strong thermal gradients in the UNM, particularly under compressive forcing. This association is much weaker in the CNM, indicating that the physical environments hosting extreme density fluctuations differ between thermal phases.

\item Extreme velocity increments show comparable enhancement of vorticity and traceless strain. Their structure is therefore associated with both rotational and shear motions rather than with a purely vorticity-dominated filamentary picture.

\item Extreme magnetic increments are the most spatially localized and preferentially occur where magnetic curl and magnetic tension are simultaneously enhanced. This association is strongly forcing dependent and is reproduced in the higher-resolution solenoidal simulation, supporting its robustness over the resolved scales.

\end{enumerate}


\begin{acknowledgments}
The Sherman Fairchild Postdoctoral Fellowship at the California Institute of Technology supports Y.H. This work used SDSC Expanse CPU and NCSA Delta CPU through allocations PHY230032, PHY230033, PHY230091, PHY230105, PHY230178, and PHY240183, from the Advanced Cyberinfrastructure Coordination Ecosystem: Services \& Support (ACCESS) program, which the National Science Foundation supports with grants \#2138259,
\#2138286, \#2138307, \#2137603, and \#2138296.
\end{acknowledgments}

\vspace{5mm}

\software{\texttt{AthenaK} \citep{2026ApJS..283...27S}, Python3 \citep{10.5555/1593511}}

\appendix

\section{Phase-conditioned plasma parameters}\label{app:plasma}
\begin{figure*}
\centering
\includegraphics[width=0.99\linewidth]{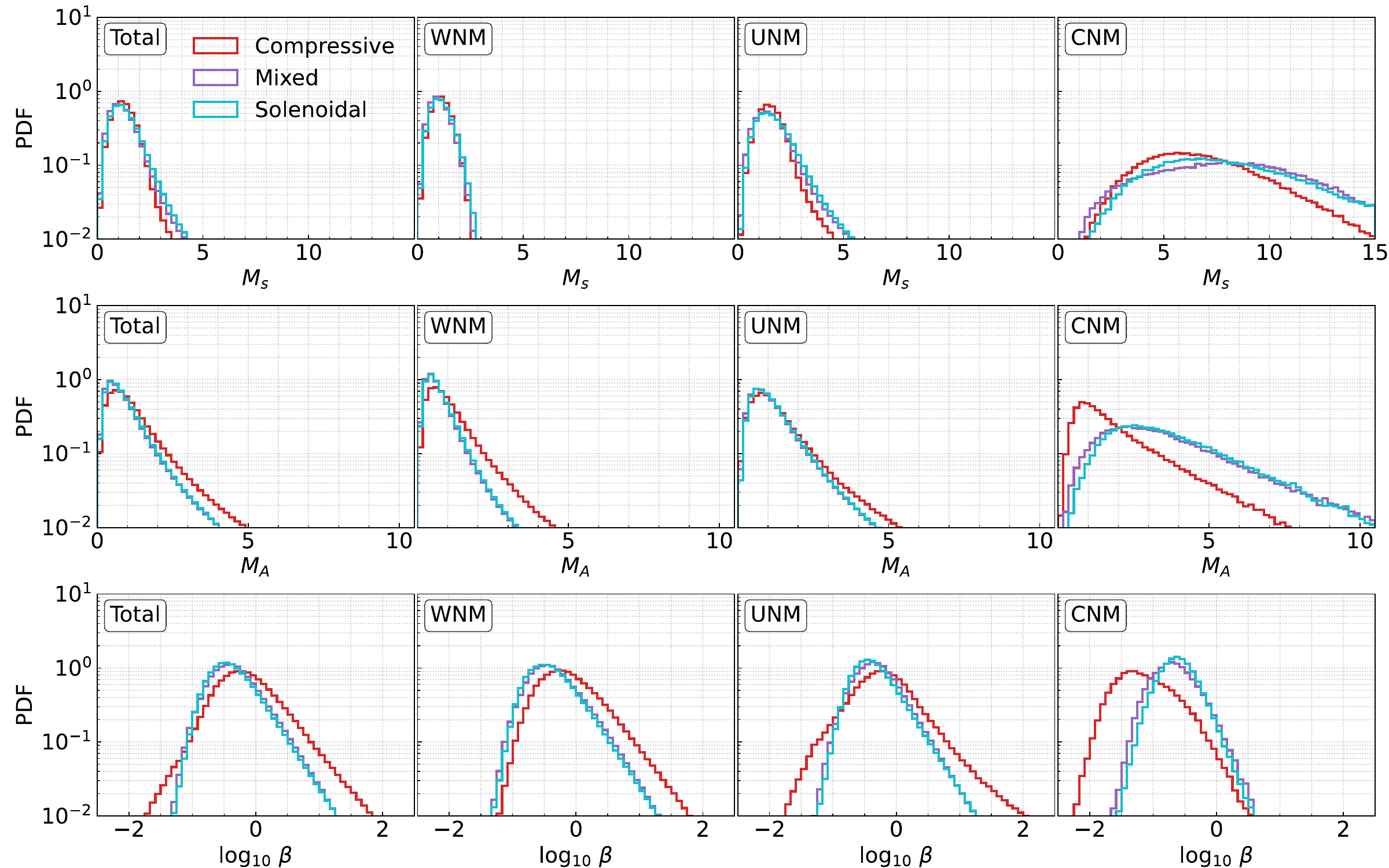}
\caption{Local \(\Ms\), \(\Ma\), and \(\log_{10}\beta\) distributions (top to bottom) for the total gas, WNM, UNM, and CNM (left to right). Red, purple, and cyan denote compressive, mixed, and solenoidal driving.}
\label{fig:machbeta}
\end{figure*}

The dimensionless parameters quoted in Table~\ref{tab:sim-properties} are computed cell by cell as
\begin{equation}
\Ms=\frac{|\boldsymbol{v}|}{c_{\rm s}}, \qquad
c_{\rm s} = \left(\frac{\gamma k_B T}{\mu m_{\rm H}}\right)^{1/2},
\label{eq:ms}
\end{equation}
\begin{equation}
\Ma=\frac{|\boldsymbol{v}|}{v_{\rm A}}, \qquad
v_{\rm A} = \frac{|\boldsymbol{B}|}{\sqrt{4\pi\rho}},
\label{eq:ma}
\end{equation}
and
\begin{equation}
\beta = \frac{P_{\rm th}}{P_{\rm mag}} = \frac{\nh k_B T}{|\boldsymbol{B}|^2/8\pi},
\label{eq:beta}
\end{equation}
with \(\gamma=5/3\) and \(\mu=1.27\).  Fig.~\ref{fig:machbeta} and Table~\ref{tab:sim-properties} show that phase-to-phase variations in \(\Ms\), \(\Ma\), and \(\beta\) exceed the forcing contrast. The WNM is transonic, the UNM is mildly supersonic, and the CNM is strongly supersonic primarily because of its low sound speed. The CNM is also magnetically dominated relative to its thermal pressure, while compressive driving produces weaker diffuse-phase fields but a stronger CNM field. These local regimes motivate the phase-conditioned analysis.

\section{Structure functions and extended self-similarity}
\label{app:sf}
\begin{figure*}
\centering
\includegraphics[width=0.99\linewidth]{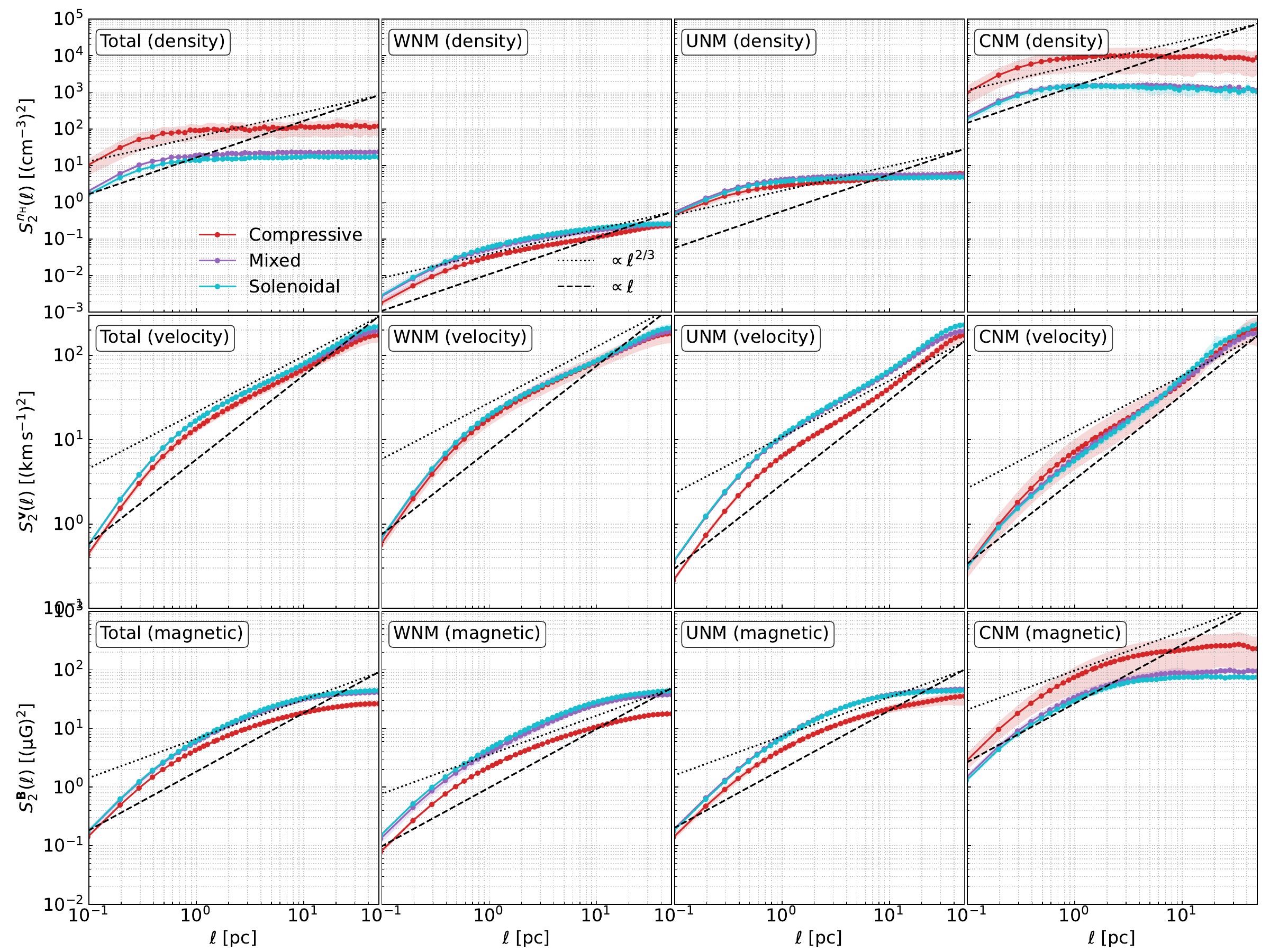}
\caption{Second-order structure functions of density, velocity, and magnetic field (top to bottom) for the total gas, WNM, UNM, and CNM (left to right). Red, purple, and cyan denote compressive, mixed, and solenoidal driving. Shaded bands show snapshot scatter; dotted and dashed lines mark \(\ell^{2/3}\) and \(\ell\).}
\label{fig:s2}
\end{figure*}

The order-\(\alpha\) structure function measures how the amplitude of field increments changes with separation:
\begin{equation}
S_\alpha(\ell) = \left\langle |\delta_\ell|^\alpha\right\rangle .
\label{eq:sf}
\end{equation}
For density, \(|\delta_\ell|\) denotes \(|\delta_\ell\nh|\); for velocity and magnetic field, it denotes the full-vector magnitude. We average over sampled positions and the three separation directions. If a power-law range exists,
\begin{equation}
S_\alpha(\ell)\propto\ell^{\zeta_\alpha},
\label{eq:sfscaling}
\end{equation}
the exponents \(\zeta_\alpha\) describe the relative scale dependence. For non-intermittent Kolmogorov scaling,
\begin{equation}
\frac{\zeta_\alpha}{\zeta_3} = \frac{\alpha}{3}.
\label{eq:k41}
\end{equation}
We show the She--L\'ev\^eque model \citep{1994PhRvL..72..336S} as a phenomenological reference,
\begin{equation}
\frac{\zeta_\alpha}{\zeta_3} = \frac{\alpha}{9} + 2\left[ 1-\left(\frac{2}{3}\right)^{\alpha/3} \right],
\label{eq:sl}
\end{equation}
although neither this model nor a single power law is assumed to describe the multiphase flow exactly. We estimate relative exponents using extended self-similarity (ESS) \citep{1993PhRvE..48...29B}:
\begin{equation}
S_\alpha(\ell) \propto \left[S_3(\ell)\right]^{\zeta_\alpha/\zeta_3},
\end{equation}
The slope of \(\log S_\alpha\) against \(\log S_3\) gives \(\zeta_\alpha/\zeta_3\), not the individual exponents. Figure~\ref{fig:s2} shows that compressive driving raises the density structure-function amplitude, especially in the CNM, while velocity and magnetic amplitudes are lower at small lags in the total, WNM, and UNM samples. Thus, forcing redistributes density and magnetic power rather than uniformly increasing it.

\begin{figure*}
\centering
\includegraphics[width=0.99\linewidth]{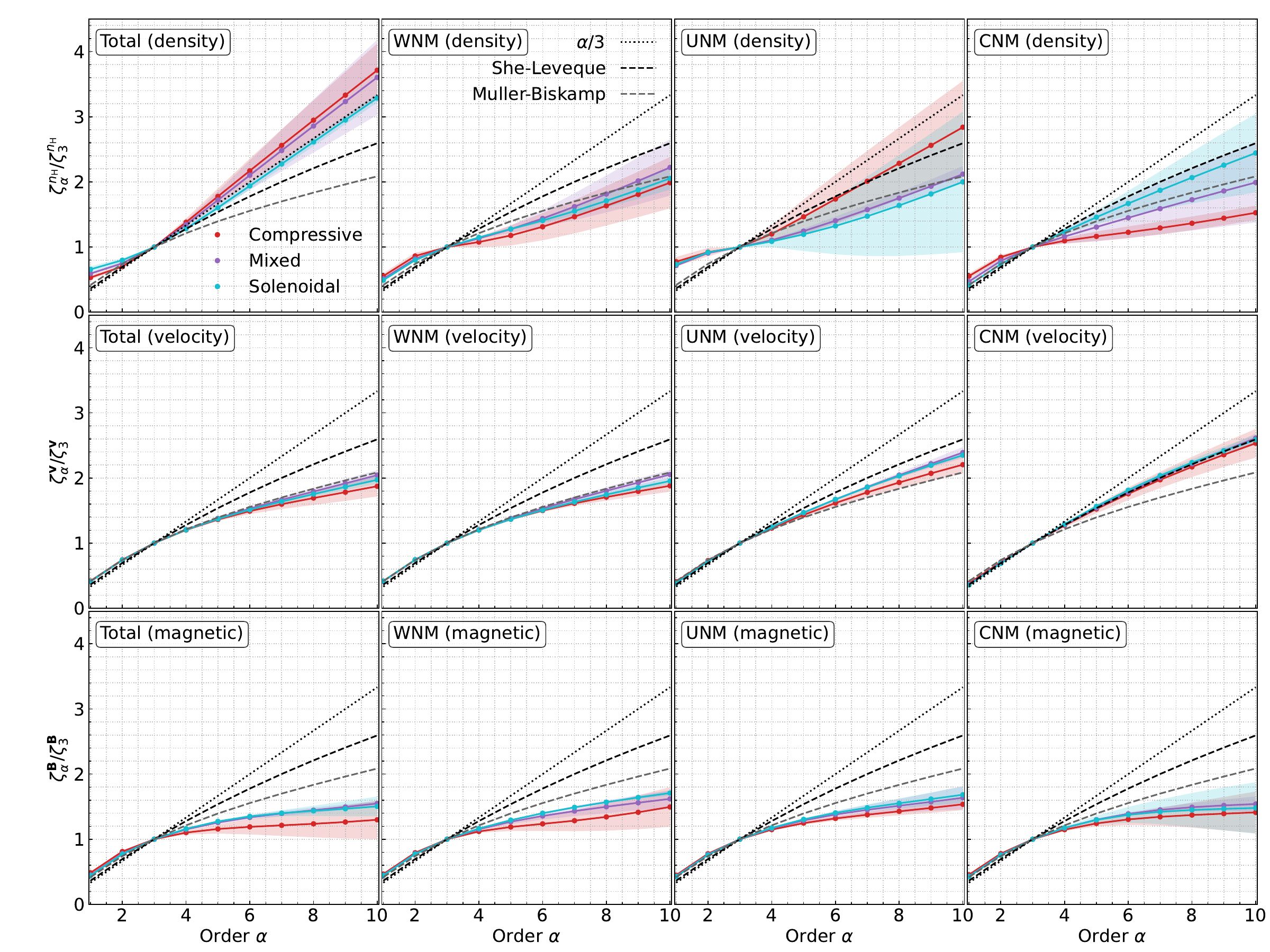}
\caption{ESS ratios \(\zeta_\alpha/\zeta_3\) for density, velocity, and magnetic field (top to bottom), separated by phase. Colors and shaded bands are as in Fig.~\ref{fig:s2}; dotted and dashed curves show \(\alpha/3\) and the She--L\'ev\^eque reference.}
\label{fig:ess}
\end{figure*}

The velocity and magnetic ESS ratios are concave and lie below \(\alpha/3\) at high order, with stronger deviations under compressive driving. The density curves do not admit the same interpretation: saturation of the density structure functions invalidates a single ESS slope over the full interval. We therefore use the ESS results only as supporting diagnostics and base the main conclusions on fixed-scale increment statistics.

\newpage
\bibliography{sample631}{}
\bibliographystyle{aasjournal}

\end{document}